\documentclass[11pt,a4paper]{article}
\pdfoutput=1

\usepackage{amsmath}
\usepackage{amsfonts}
\usepackage{amsthm}
\usepackage{amssymb}
\usepackage{amscd}
\usepackage[british]{babel}
\usepackage{graphicx}
\usepackage{psfrag}
\usepackage{epsfig}
\usepackage{rotating}
\usepackage{times}
\usepackage{color}

\theoremstyle{plain}

\newcommand{\boxend}{\flushright{$\Box$}}

\newcommand{\R}{{\mathbb R}}               

\renewcommand{\tilde}{\widetilde}

\begin{document}

\title{Viability of the Matter Bounce Scenario }

\author{Jaume de Haro$^{}$\footnote{E-mail: jaime.haro@upc.edu}
and Jaume Amor\'os$^{}$\footnote{E-mail: jaume.amoros@upc.edu}}

\maketitle

{Departament de Matem\`atica Aplicada I, Universitat
Polit\`ecnica de Catalunya, Diagonal 647, 08028 Barcelona, Spain}

\thispagestyle{empty}

\begin{abstract}
It is shown that teleparallel $F({\mathcal T})$  theories of 
gravity combined with Loop Quantum Cosmology support a Matter Bounce Scenario  which is an alternative to 
the inflation scenario in the Big Bang paradigm. It is checked that
these bouncing models 
provide theoretical data that fits well with the current observational data, allowing the viability of the Matter Bounce Scenario.
\end{abstract}

\section{Introduction}
It is well-known that inflation suffers from several problems (see \cite{brandenberger} for a review about these problems), like
 the initial singularity which  is usually not addressed,
   or the fine-tuning of the degree of flatness required for the potential in order to achieve successful inflation \cite{ffg91}.
   
 In order to avoid these problems,
an alternative scenario to the inflationary paradigm, called {\em Matter Bounce Scenario} (MBS), has been developed in order to
explain the evolution of our Universe (see \cite{cai}). Essentially, it depicts at very early times a matter dominated Universe in a
contracting phase, that evolves towards the bounce and afterwards enters 
an expanding phase. This model, like inflation, solves the horizon problem that appears in General Relativity (GR) and improves the flatness problem
in GR (where spatial flatness is an unstable fixed point and  fine tuning of initial conditions is required),
because
the contribution of the spatial curvature decreases in
the contracting phase at the same rate as it increases in the expanding one (see for instance \cite{brandenberger1}).  
   
The aim of our work is to construct viable bouncing cosmologies where the matter part of the Lagrangian is composed of a single scalar field and, therefore, have to 
 go beyond  General Relativity, since GR forbids  bounces when one deals 
 with a single field. Hence,  theories such as  holonomy corrected Loop Quantum Cosmology (LQC) \cite{wilson}, where a big bounce appears owing 
 to the discrete structure of space-time \cite{ashtekar} or teleparalellism \cite{Haro1} must be taken into account.
  When dealing with these theories, in order to obtain a theoretical value of the
 spectral index and its running that may fit well with current experimental data, a quasi-matter dominated regime in 
 the contracting phase 
 termed by the condition $\left|w\equiv\frac{P}{\rho}\right|\ll 1$, where $P$ and $\rho$ are respectively the pressure and the energy density of the Universe,
 has to be introduced \cite{eho}.
 
 Since in
Matter Bounce Scenario the number of e-folds
before the end of the quasi-matter domination regime
can be relatively small, the horizon problem does not exist in bouncing cosmologies and the flatness problem is neutralized {}{\cite{brandenberger1}}. This argues for the
viability of such models, making it possible that for certain matter bounce scenarios the forecast values of the spectral index and of the running parameter agree well with the most accurate current observations.
 
In contrast, in slow roll inflation one must consider the running of the spectral index
corresponding
to $N$ e-folds before the end of the inflation, which
in general, is  of the order of $N^{-2}$.
This
value turns out to be very small, when one substitutes for $N$ the minimum number of e-folds which are needed to solve the horizon and flatness problem in inflationary cosmology 
(the usual accepted value is $N> 50$), 
as compared with its corresponding observational value
$-0.0134\pm 0.009$ coming from the most recent Planck data \cite{Ade}. This shows that these slow roll models are less favored by observations. 

\vspace{0.25cm}

The units used in the paper are: $\hbar=c=8\pi G=1$.

\section{$F({\mathcal T})$ gravity in flat FLRW geometry}

Teleparallel theories are based in the {\em Weitzenb\"ock space-time}. 
This space is $\R^4$, with a Lorentz metric, in which a global, orthonormal basis of its tangent bundle given by 
four vector fields $\{ e_i \}$ has been selected, that is, they satisfy
$g(e_i, e_j)= \eta_{ij}$ with $\eta=\text{diag}\,(-1,1,1,1)$. The 
Weitzenb\"ock 
connection $\nabla$ is defined by imposing that the basis vectors $e_i$ be absolutely parallel, i.e. that $\nabla e_i=0$.

The Weitzenb\"ock connection is compatible with the metric $g$, 
and it has zero curvature because of the global parallel transport
defined by the basis $\{ e_i \}$. The information of the 
Weitzenb\"ock connection is carried by its torsion,
and its basic invariant is the {\em scalar torsion} ${\mathcal T}$.
The connection, and its torsion, depend on the choice of orthonormal
basis $\{ e_i \}$, but if one adopts the flat Friedmann-Lema\^{\i}tre-Robertson-Walker (FLRW) metric and selects as orthonormal basis $\{ e_0= \partial_0,
e_1= \frac{1}{a} \partial_1, e_2= \frac{1}{a} \partial_2, 
e_3= \frac{1}{a} \partial_3 \}$,
then the scalar torsion is 
\begin{eqnarray}\label{eq:-6H2}
{\mathcal T}=-6H^2 \; ,
\end{eqnarray}
where $H=\frac{\dot{a}}{a}$ is the Hubble parameter,
 and this identity is invariant with respect to local Lorentz transformations that only depend on the time, i.e. of the form
$\tilde{e_i}= \Lambda^k_i(t) e_k$ (see \cite{Bengochea:2008gz,ha}).

\bigskip

With the above choice of orthonormal fields, the Lagrangian of
the $F({\mathcal T})$ theory of gravity is
\begin{eqnarray}\label{eq:lagr}
{\mathcal L}_{\mathcal T}={\mathcal V}(F({\mathcal T})+{\mathcal L}_M),
\end{eqnarray}
where ${\mathcal V}=a^3$ is the volume of the Universe, 
and ${\mathcal L}_M$ is the matter Lagrangian density.

The Hamiltonian of the system is 
\begin{eqnarray}\label{eq:ham}
{\mathcal H}_{\mathcal T}= \left(2{\mathcal T}\frac{dF({\mathcal T})}{d{\mathcal T}}-F({\mathcal T})  +\rho \right){\mathcal V} \, ,
\end{eqnarray}
where ${\rho}$ is the energy density. Imposing the Hamiltonian constrain ${\mathcal H}_{\mathcal T}=0$ leads to the 
modified Friedmann equation
\begin{eqnarray}\label{eq:friedmann}
\rho=-2 \frac{dF({\mathcal T})}{d{\mathcal T}}{\mathcal T}+F({\mathcal T})\equiv G({\mathcal T})
\end{eqnarray}
which, as $\mathcal T=-6H^2$, defines a curve in the plane 
$(H,\rho)$.

Equation (\ref{eq:friedmann}) may be inverted, so
a curve of the form $\rho=G({\mathcal T})$ defines an $F({\mathcal T})$ theory with
\begin{eqnarray}\label{eq:F(T)}
F({\mathcal T})=-\frac{\sqrt{-{\mathcal T}}}{2}\int \frac{G({\mathcal T})}{{\mathcal T}\sqrt{-{\mathcal T}}}d{\mathcal T}.
\end{eqnarray}

\bigskip

To produce a cyclically evolving Universe, let us take the $F({\mathcal T})$ theory arising from the ellipse that defines the holonomy corrected Friedmann equation in Loop Quantum
Cosmology 
\begin{eqnarray}
H^2=\frac{\rho}{3}\left(1-\frac{\rho}{\rho_c}\right),
\end{eqnarray}
where $\rho_c$ is the so-called {\it critical density}.

To obtain a parametrization of the form $\rho=G({\mathcal T})$,
the curve has to be split in two branches
\begin{eqnarray}\label{eq:Gpm}
\rho=G_{\pm}({\mathcal T})=\frac{\rho_c}{2}\left(1\pm\sqrt{1+\frac{2{\mathcal T}}{\rho_c}}\right),
\end{eqnarray}
where the branch $\rho=G_-({\mathcal T})$ corresponds to $\dot{H}<0$
and $\rho=G_+({\mathcal T})$ is the branch with $\dot{H}>0$.
Applying Eq. (\ref{eq:F(T)}) to these branches produces the model
(\cite{aho13,aho14,as})
\begin{eqnarray}\label{eq:Fpm}
F_{\pm}({\mathcal T})=\pm\sqrt{-\frac{{\mathcal T}\rho_c}{2}}\arcsin\left(\sqrt{-\frac{2{\mathcal T}}{\rho_c}}\right)+G_{\pm}({\mathcal T}).
\end{eqnarray}

\section{Matter Bounce Scenario}
Matter Bounce Scenarios (see \cite{cai} for a recent review) are essentially characterized by the Universe being nearly matter dominated at very early times in the contracting phase (to
obtain an approximately scale invariant power spectrum)
and  evolving 
towards a bounce where all the parts of the Universe become in causal contact \cite{aho13}, solving the horizon problem, to enter into a expanding regime, where it matches 
the behavior of the standard hot Friedmann Universe. They constitute an alternative to the inflationary paradigm.

According to the current observational data, in order to obtain a  viable  MBS model, the bouncing model has
to satisfy  some conditions that we have summarized  as  follows:
\begin{enumerate}
 \item
 The latest Planck data constrain the value of the spectral index for scalar perturbations and its running, namely $n_s$ and $\alpha_s$,
to $0.9603\pm 0.0073$ and $-0.0134\pm 0.009$ respectively \cite{Ade}.
The analysis of these parameters provided by Planck  makes no slow roll
approximation (in fact, the determination of cosmological parameters from the first year WMAP observations was done considering the $\Lambda$CDM model \cite{verde}), which means that 
the parameters $n_s$ and $\alpha_s$ could be used to test bouncing models. On the other hand,
it is well-known that the  ways to obtain a nearly scale invariant power spectrum of perturbations with running are either a quasi de Sitter phase in the expanding phase or a nearly
matter domination phase at early times, in the contracting phase \cite{w99}. Then, since
for the MBS one has $n_s=1$,  if one wants to improve the model to match correctly with this observational data, one has to consider, at early times in the 
contracting phase,
a  quasi-matter domination period characterized by the condition $\left| w\equiv \frac{P}{\rho}\right|\ll 1$, being $P$ and $\rho$ the pressure and the energy density of the Universe.

\item The Universe has to reheat creating light particles that will thermalize matching with a hot Friedmann Universe. 
 Reheating could  be produced due to the gravitational particle creation in an
expanding Universe \cite{geometric}. In this case, an abrupt phase transition (a non adiabatic transition) is needed in order to obtain sufficient particle creation that thermalizes 
producing a reheating temperature that fits well with current observations.
This method was used in the context of inflation in \cite{ford,peebles}, where a sudden phase transition from a quasi de Sitter phase
to a radiation domination or a quintessence phase was assumed in the expanding regime. It is shown in \cite{Quintin} that
gravitational particle production  could be applied to the MBS, assuming a phase transition from the matter domination
to an ekpyrotic phase in the contracting regime, and obtaining a reheating temperature compatible with current data.

\item  Studies of distant type Ia
supernovae (\cite{p99} and others)  provide strong evidence that our
Universe is expanding in an accelerating way. A viable model must take into account this current acceleration, which could be incorporated, in the simplest case, with a cosmological
constant, or by quintessence models \cite{quintessence}.
There are other ways to implement the current cosmic acceleration, for example using $F({\mathcal R})$ gravity (see for instance \cite{odintsov}), but the current models
that provide this behavior are very complicated, and the main objective  in MBS is to present the simplest viable models.

\item 
The data of the seven-year survey WMAP (\cite{Larson})
constrains the value of the power spectrum for
scalar perturbations to be ${\mathcal P}_S(k)\cong 2\times 10^{-9}$.
The numerical  results (analytical ones will be impossible to obtain) calculated with bouncing models have to match with  that experimental data.

\item
The constrain of the tensor/scalar ratio provided by WMAP and Planck projects (${r}\leq 0.11$) is obtained indirectly assuming the  {\it consistency} slow roll relation
${r}=16\epsilon$ (where $\epsilon=-\frac{\dot{H}}{H^2}\cong \frac{1}{2}\left(\frac{V_{\varphi}}{V} \right)^2$ is the main slow roll parameter) \cite{Peiris}, because gravitational waves 
are not longer detected by those projects. This means that, the slow roll inflationary models must satisfy this constrain, but
not the bouncing ones, where there is not any consistency relation. 
This point is very important because some very complicated mechanisms are sometimes implemented to the MBS  in order to enhance the
power spectrum of scalar perturbation to achieve the observational bound provided by Planck \cite{Cai1}.
In fact, in matter bounce scenario, to check  if the models provide a viable value of the
tensor/scalar ratio, first of all gravitational waves must be clearly detected in order to determine the observed value of this ratio. The authors
hope that more accurate unified Planck-BICEP2 data (the B2P collaboration), which is going to be issued soon, may 
adress this point. In contrast, as we have pointed out in (i),  the spectral index of scalar perturbations
and its running could be calculated independently of the theory, which means that in order to check bouncing models, while in the absence of evidence  of gravitational waves, one
has to work in the space $(n_s,\alpha_s)$.

\end{enumerate}

\section{Perturbations in Matter Bounce Scenario}

The Mukhanov-Sasaki equations (see \cite{sasaki} for a deduction of these equations in GR) for $F({\mathcal T})$ gravity and LQC are given by \cite{Cai,Cailleteau}

\begin{eqnarray}
\zeta_{S(T)}''-{c}^2_{s}\nabla^2 \zeta_{S(T)}+\frac{Z_{S(T)}'}{Z_{S(T)}}\zeta'_{S(T)}=0,
\end{eqnarray}
 where $\zeta_{S}$ and $\zeta_{T}$ denote the amplitude for scalar and tensor perturbations.
 
In   $F({\mathcal T})$ gravity one has 
\begin{eqnarray}
Z_S=\frac{a^2{|\Omega|}\dot{{\varphi}}^2}{{c}^2_{s}{H^2}},\quad
Z_T=\frac{a^2c^2_s}{{|\Omega|}},
\quad
c_s^2=|\Omega|\frac{
\arcsin\left(2\sqrt{\frac{3}{\rho_c}}H\right)}{2\sqrt{\frac{3}{\rho_c}}H}, \quad \mbox{with} \quad \Omega=1-\frac{2\rho}{\rho_c}.\end{eqnarray}

In contrast, for LQC,
\begin{eqnarray}
 Z_S=\frac{a^2\dot{{\varphi}}^2}{{H}^2},\quad
Z_T=\frac{a^2}{{\Omega}},\quad c^2_s=\Omega.
\end{eqnarray}

The power spectrum for scalar perturbations is given by \cite{ha14}
\begin{eqnarray}
{\mathcal P}_{S}(k)=\frac{3\rho_c^2}{\rho_{pl}}
 \left|\int_{-\infty}^{\infty}{Z^{-1}_{S}(\eta)}d\eta\right|^2,
\end{eqnarray}
where, in order to obtain this formula, the scale factor $a(t)\cong (\frac{4}{3}\rho_c t^2)^{1/3}$ at early times has been used. In the particular case of an
exactly matter dominated universe during all the background evolution, i.e., when  $a(t)=(\frac{4}{3}\rho_c t^2+1)^{1/3}$  for teleparalell $F({\mathcal T})$  gravity one has
$ {\mathcal P}_{S}(k)
=\frac{16}{9}\frac{\rho_c}{\rho_{pl}}{\mathcal C }^2, 
$ \cite{h13}
where ${\mathcal C }=1-\frac{1}{3^2}+\frac{1}{5^2}-...=  0.915965...$ is the  Catalan's constant, and for holonomy corrected LQC 
 ${\mathcal P}_{S}(k)
=\frac{\pi^2}{9}\frac{\rho_c}{\rho_{pl}} 
$ \cite{w13}.

The ratio of tensor to scalar perturbations in MBS is given by
\begin{eqnarray}
r
=\frac{8}{3}\left(\frac{\int_{-\infty}^{\infty}{Z^{-1}_T(\eta)}d\eta}{\int_{-\infty}^{\infty}{Z^{-1}_S(\eta)}d\eta}\right)^2,
\end{eqnarray}
where the factor 8 appears due to the two polarizations of the gravitational waves and to the renormalization with respect to a canonical field \cite{langlois}.

The spectral index for scalar perturbations and its running are calculated in \cite{eho} given
\begin{eqnarray}
 n_s-1=12w,\quad \alpha_s=-48\delta^2,
\end{eqnarray}
where the parameters $w$ and $\delta^2$, calculated in the quasi-matter domination,  as a functions of the potential are
\begin{eqnarray}
 w
 \cong \frac{1}{3}\left(\frac{V_{\varphi}}{V} \right)^2-1,\quad {\delta}^2
 \cong  -\left(\frac{V_{\varphi}}{V} \right)_{\varphi}.
\end{eqnarray}

\subsection{Comparison with observational data in the plane $(n_s,\alpha_s)$}
In slow-roll inflation, 
for the general models (monomial, natural, hilltop and plateau potentials),  $1-n_s$ is of the order $N^{-1}$, while the running parameter is of order $N^{-2}$ and, consequently,   one has
$\alpha_s\sim (1-n_s)^2$, which in most cases is incompatible with Planck and WMAP data, because the observed value of the running is not small enough
\cite{running, cai1}. 

Thus, the observation of a large negative running implies that
any inflationary phase requires multiple fields or the breakdown of slow roll. Following  this second path, in \cite{cai1} the authors consider
the break of the slow-roll approximation  for a short while, due to the
inclusion of a quickly oscillating term in the potential. As a consequence,  the theoretical value of the running parameter gets larger and could match well with observational data. 

In contrast, in MBS the situation is completely different. For example, in \cite{eho} dealing with a perfect fluid
whose Equation of State (EoS) is parametrized by the number of e-folds before the end of the quasi-matter domination period, namely $N$, the authors have shown that the theoretical values of the spectral index of scalar perturbations and its running fit well with their corresponding observational data. To be more precise,
for the
 EoS ${P}=\frac{\beta}{(N+1)^{\alpha}}\rho$, ($\alpha>0$, $\beta<0$) 
the following relation
\begin{eqnarray}\alpha_s=\frac{2\alpha}{N+1}(n_s-1)\end{eqnarray}
is obtained, 
which is perfectly compatible with the experimental data.
In fact, for instance, if one takes $\alpha=2$ and $N=12$ (note that in bouncing cosmologies a large number of e-folds is {\it not} required, because the horizon problem does not exist, since
at the bounce all  parts of the Universe are already in causal contact,
and also the flatness problem gets improved \cite{brandenberger1}),
one obtains, for $n_s=0.9603\pm 0.0073$, the following value for the running parameter: $\alpha_s=-0.0122\pm 0.0022$, which is
compatible with the Planck data. Effectively, for these values of $\alpha$ and $N$ one gets $n_s-1=\frac{12}{13^2}\beta\cong 0.071\beta$, which is indeed compatible with its
observed value, by choosing $\beta\cong-\frac{1}{2}$.


\section*{Acknowledgements}

The authors would like to thank Professor Sergei D. Odintsov for his valuable and useful comments.
This investigation has been
supported in part by MINECO (Spain), project MTM2011-27739-C04-01, MTM2012-38122-C03-01.



\begin{thebibliography}{99}

\bibitem{brandenberger}
R.H. Brandenberger,
(2012)
  [arXiv:astro-ph/1206.4196].\\
  R.H. Brandenberger, Int. J. Mod. Phys. Conf. Ser {\bf 01}, 67 (2008) [arXiv:0902.4731].
\bibitem{ffg91}
F.C. Adams, K. Freese and A.H. Guth, Phys. Rev. {\bf D43}, 965 (1991).
\bibitem{cai}
Yifu Cai, SCIENCE CHINA: Phys. Mech. Astr. {\bf 57},  1414 (2014) [arXiv: 1405.1369].
\bibitem{brandenberger1}
R.H. Brandenberger,
(2012)
  [arXiv:1204.6108].
  
  \bibitem{wilson}
Yifu Cai and E. Wilson-Ewing, JCAP {\bf 03} 026, (2014) [arXiv:1402.3009]. \\
M. Bojowald and G.M. Hossain, Phys. Rev. {\bf D77}, 023508 (2008) [arXiv: 0709.2365].\\
T. Cailleteau, J. Mielzczarek, A. Barrau and J. Grain, Class. Quant. Grav. {\bf 29}, 095010 (2012) [arXiv:111.3535].

\bibitem{ashtekar}
A. Ashtekar and P. Singh, Class. Quant. Grav. {\bf 28} 23001, (2011) [arXiv:1108.0893].\\
P. Singh,  Class. Quant. Grav. {\bf 26} 125005, (2009)
  [arXiv:0901.2750]. 
  
    \bibitem{Haro1}
 K. Bamba, J. de Haro and S.D. Odintsov, JCAP 02
{\bf 008}, (2013) [arXiv:1211.2968].
 \bibitem{eho}
  E. Elizalde, J. Haro and S.D. Odintsov (2014) [arXiv:1411.3475].
   
  \bibitem{Ade}
P.A.R. Ade et al., Astronomy and Astrophysics {\bf 571},  A22 (2014)  [arXiv:1303.5082].
  
\bibitem{Bengochea:2008gz}
  G.~R.~Bengochea and R.~Ferraro,
  Phys.\ Rev.\  D {\bf 79}, 124019 (2009)
  [arXiv:0812.1205].

\bibitem{ha}
J. de Haro and J. Amor\'os, Phys. Rev. Lett. {\bf 110},  071104  (2013) [arXiv:1211.5336].

\bibitem{aho13}  J. Amor\'os, J. de Haro and S.D. Odintsov, Phys. Rev. {\bf D87}, 104037 (2013) [arXiv:1305.2344].

\bibitem{aho14}
  J. Amor\'os,  J. de Haro and S.D. Odintsov, Phys. Rev. {\bf D89}, 104010 (2014)   [arXiv:1402.3071].

\bibitem{as}
A. Ashtekar and P. Singh,  Class. Quantum Grav. {\bf 28}, 213001 (2011) [arXiv: 1108.0893].

\bibitem{verde}
D.N. Spergel et al.,  Astrophys.J.Suppl. {\bf 148}, 175 (2003)  [arXiv:0302209].
 \\
 L. Verde et al., Astrophys.J.Suppl. {\bf 148}, 195 (2003) [arXiv:0302218].

 
\bibitem{w99}
  D. Wands, Phys Rev. {\bf D 60}, 023507 (1999) [arXiv:9809062].

\bibitem{geometric}
A.A. Grib, S.G. Mamayev  and V.M. Mostepanenko, Gen. Rel. Grav. {\bf 7}, 535 (1976).   \\
 A.A. Grib, S.G. Mamayev  and V.M. Mostepanenko, {\it Vacuum Quantum Effects in Strong Fields}, St Petersburg:
Friedmann Laboratory Publishing (1994).\\
L. Parker, Phys. Rev. {\bf 183}, 1057 (1969).
\bibitem{ford}
 L.H. Ford,  Phys. Rev. {\bf D35}, 2955 (1985).
\bibitem{peebles}
 P.J.E. Peebles and A. Vilenkin, Phys. Rev. {\bf D59}, 063505 (1999) [arXiv:9810509].
\bibitem{Quintin}
 J. Quintin, Yifu Cai and R. Brandenberger, Phys. Rev. {\bf D90}, 063507 (2014)[arXiv:1406.6049]
\bibitem{p99}  S. Perlmutter et al.,  Astrophys. J. {\bf  517}, 565 (1999) [arXiv:9812133].\\
  A.G. Riess et al.,  Astron. J. {\bf  116}, 1009 (1999) [arXiv:9805201].
 \bibitem{quintessence}
 R.R. Caldwell, R. Dave and P.J. Steinhardt, Phys. Rev. Lett. {\bf 80}, 1582 (1988).\\
 I. Zlatev, L. Wang and P.J. Steinhardt,    Phys. Rev. Lett.{\bf 82}, 896 (1999)  [arXiv:9807002].
 
\bibitem{odintsov} 
 S. Nojiri and S.D. Odintsov,  Int.J.Geom.Meth.Mod.Phys. {\bf 4}, 115 (2007) [arXiv:0601213].\\
 S. Nojiri and S.D. Odintsov,           Phys.Rept.{\bf 505}, 59 (2011) [arXiv:1011.0544].

\bibitem{Larson}
D. Larson et al., Astrophys. J. Suppl. {\bf 192}, 16 (2011)
  [arXiv:1001.4635].
\bibitem{Peiris}
H. Peiris and R. Easther, JCAP {\bf 0607}, 002 (2006) [arXiv:0603587].\\
S.M. Leach, A.R. Liddle, J. Martin and D.J. Schwarz, Phys.Rev. {\bf D66},  023515 (2002) [arXiv:0202094].
S.M. Leach and  A.R. Liddle, Mon.Not.Roy.Astron.Soc. {\bf 341},   1151 (2003) [arXiv:0207213].

\bibitem{Cai1}
  Yifu Cai, R. Brandenberger and X. Zhang, JCAP {\bf 1103},
003 (2011) [arXiv:1101.0822].
\\
 Yifu Cai, D. A. Easson and R. Brandenberger, JCAP {\bf 1208},
020 (2012) [arXiv:1206.2382].

\bibitem{sasaki}
  V.F. Mukhanov, JETP Lett. {\bf 41}, 493 (1985).\\
  M. Sasaki, Prog. Theor. Phys. {\bf 76}, 1036 (1986).

\bibitem{Cai}
 Yifu Cai,  S-H. Chen, J.B. Dent, S. Dutta and E.N. Saridakis, Class.Quant.Grav. {\bf 28},  215011 (2011) [arXiv:1104.4349].


\bibitem{Cailleteau}
T. Cailleteau,  A. Barrau, J. Grain and F. Vidotto,  PRD {\bf 86},   087301 (2012)  [arXiv:1206.6736].
\bibitem{ha14}
 J. de Haro and J. Amor\'os,  JCAP {\bf 08}, 025(2014)  [arXiv:1403.6396].
 \bibitem{h13}
J. Haro, JCAP {\bf 11}, 068 (2013) [arXiv:1309.0352].
 \bibitem{w13}
  E. Wilson-Ewing, JCAP {\bf 03}, 026 (2013) [arXiv:1211.6269].
\bibitem{langlois}
D. Langlois, (2010) [arXiv:1001.5259].
\bibitem{running}
R. Easther and H. Peiris, JCAP {\bf 010}, 0609 (2006) [arXiv:0604214].
\bibitem{cai1}
Y. Wan, S. Li, M. Li, T. Qiu, Yifu Cai,
and X. Zhang Phys. Rev. {\bf D90}, 023537 (2014) [arXiv:1405.2784].




\end{thebibliography}
\end{document}